\documentclass[citeautoscript,floatfix,aps,prb,twocolumn,
      superscriptaddress]{revtex4-2}

\usepackage{amsmath}
\usepackage{graphicx}
\usepackage{tabularx}
\usepackage{float}
\usepackage{epstopdf}
\usepackage{natbib}
\usepackage{array}
\usepackage{braket}
\usepackage[usenames,dvipsnames]{color}
\usepackage{dcolumn}
\usepackage{bm}
\usepackage{soul}  
\usepackage{changes}
\usepackage{hyperref}
\hypersetup{colorlinks=true, linkcolor=blue, filecolor=blue, urlcolor=blue}
\begin{document}
\title{Quantum embedding study of strain and charge induced Stark effects on the NV$^{-}$ center in diamond}
\author{Gabriel I. L\'{o}pez-Morales}
\affiliation{Department of Physics, City College of the City University of New York, New York, New York 10031, USA}
\author{Joanna M. Zajac}
\affiliation{Instrumentation Division, Brookhaven National Laboratory, Bldg. 535 P.O. Box 5000, Upton, New York 11973-5000, USA}
\author{Johannes Flick}
\affiliation{Department of Physics, City College of the City University of New York, New York, New York 10031, USA}
\affiliation{Center for Computational Quantum Physics, Flatiron Institute, 162 5th Avenue, New York, New York 10010, USA}
\affiliation{The Graduate Center of the City University of New York, New York, New York 10016, USA}
\author{Carlos A. Meriles}
\affiliation{Department of Physics, City College of the City University of New York, New York, New York 10031, USA}
\affiliation{The Graduate Center of the City University of New York, New York, New York 10016, USA}
\author{Cyrus E. Dreyer}
\affiliation{Department of Physics and Astronomy, Stony Brook University, Stony Brook, New York, 11794-3800, USA}
\affiliation{Center for Computational Quantum Physics, Flatiron Institute, 162 5th Avenue, New York, New York 10010, USA}

\date{\today}

\begin{abstract}
The NV$^{-}$ color center in diamond has been demonstrated as a powerful nanosensor for quantum metrology due to the sensitivity of its optical and spin properties to external electric, magnetic, and strain fields. In view of these applications, we use quantum embedding to derive a many-body description of strain and charge induced Stark effects on the NV$^{-}$ center. We quantify how strain longitudinal to the axis of NV$^{-}$ shifts the excited states in energy, while strain with a component transverse to the NV$^{-}$ axis splits the degeneracies of the $^{3}E$ and $^{1}E$ states. The largest effects are for the optically relevant $^{3}E$ manifold, which splits into $E_{x}$ and $E_{y}$ with transverse strain. From these responses we extract strain susceptibilities for the $E_{x/y}$ states within the quasi-linear regime. Additionally, we study the many-body dipole matrix elements of the NV$^{-}$ and find a permanent dipole 1.64 D at zero strain, which is somewhat smaller than that obtained from recent density functional theory calculations. We also determine the transition dipole between the $E_{x}$ and $E_{y}$ and how it evolves with strain.

\end{abstract}
\maketitle
\section{Introduction \label{sec:introduction}}
The negatively-charged nitrogen vacancy (NV$^{-}$) center in diamond is the most well-developed platform for realizing color-center based quantum and nanoscale technologies. As its name suggest, the NV$^{-}$ is formed through a combination of a carbon vacancy and an adjacent nitrogen substitution, with an excess electronic charge that typically comes from a nearby donor defect in the diamond lattice. Extensive experimental and theoretical work on its mechanisms of formation, charge-state stability and dynamics, as well as its interesting spin properties, has led to a well-founded understanding about the overall electronic structure of this color center in diamond ~\cite{Rogers2009, Doherty2011, Doherty2012, Doherty2013, Deák2014, Thiering2018, Manson2018, Razinkovas2021}.

In particular, nanoscale sensing constitutes an area of applications where the optoelectronic properties of the NV$^{-}$ have been exploited with great success ~\cite{Schirhagl2014, Balasubramanian2014, Zhang2021}. High-precision temperature sensing, magnetometry-based bio-imaging, and noise spectroscopy at the nanoscale, are all recent applications based on optical and microwave manipulation of the excited states of the NV$^{-}$ center through its spin-selective transitions ~\cite{Neumann2013, Laraoui2015, Zhang2021, Rovny2022, Monge2023}. The NV$^{-}$ is also useful in sensing local electric fields at the nanoscale with high precision ~\cite{Tamarat2006, Dolde2011, Bassett2011, Mittiga2018, Bian2021, Ji2024, Delord2024}. The lack of inversion symmetry in this color center facilitates the emergence of static electric dipoles for some of its electronic states ~\cite{Maze2011, Doherty2013, Gali2019}, which interact with the external (e.g., local or applied) electric fields such that it results in spectral shifts of  optical transitions. This phenomenon is broadly known as the electric Stark effect. The presence of local or macroscopically-applied strain fields has somewhat equivalent effects on the NV$^{-}$; i.e., they introduce strain-induced Stark shifts or splitting of the optical lines ~\cite{Maze2011, Doherty2013, Olivero2013, Gali2019, McCullian2022, McCullian2024}. Altogether, these Stark effects in the NV$^{-}$ could be further harnessed to implement novel NV-based spectroscopy and sensing protocols~\cite{Olivero2013, McCullian2022, Trusheim2016, Kehayias2019}, including for high-energy detectors \cite{BudnikPRB18, LukinPRD17, Kirkpatrick2023, WalsworthAVSQuantumSci22, WalsworthOptLett17}

For use of NV$^{-}$ as a nanosensor, it is crucial to have a quantitative understanding of the strain and electric field induced Stark shifts. The first experimental study to estimate the electric-field susceptibility (permanent dipole) of the NV$^{-}$ provided 1.2--1.5 D ~\cite{Tamarat2006} ~\footnote{Note that to compare with our calculations described below (which do not involve macroscopic fields), the relevant results from, e.g., Ref.~\citenum{Tamarat2006} are those where the screening of macroscopic fields by the crystal are accounted for to convert to a response to the local field felt by the NV$^-$}; more recently, this number was estimated to have longitudinal and transverse components of 2.82 and 3.63 D, respectively ~\cite{Ji2024}. Studies employing the modern theory of polarization through density functional theory (DFT) have found a permanent dipole of 4.33 D ~\cite{Gali2019}, compared to 2.23--2.68 D obtained through direct computation of the electric-field Stark shift in the NV$^{-}$ through DFT ~\cite{Alaerts2024}. A calculation based on the piezoelectric properties of NV$^-$, on the other hand, finds 0.79 D ~\cite{Maze2011}. Other works employing few-level Hamiltonians and molecular-orbital approaches to treat the properties of the NV$^{-}$ under strain provided useful insights on the strain responses, but did not provide explicit numbers for the dipole couplings ~\cite{Rogers2009, Doherty2011}. Although some of these theoretical results are relatively close to experimental observations and provide valuable information about Stark effects on the NV$^{-}$, further work is required to fully quantify its dipole couplings and understand the physics underlying the evolution of the optical excited states of this color center under minute external (e.g., strain, electric) fields. 

In particular, this defect center features a non-negligible degree of electronic correlations within its set of $sp^{3}$-hybridized localized electrons, two of which are unpaired. It has become increasingly evident that an adequate treatment of such correlations is required to quantitatively describe the excited-state properties of the NV$^{-}$ ~\cite{Bockstedte2018, Ma2021, Bhandari2021, Muechler2022, Chen2023}. In these regards, a full $\textit{ab-initio}$ description of the many-body physics of this defect center, most of which is central for applications such as nanoscale sensing, is highly desirable and constitutes an area of ongoing research efforts ~\cite{Bockstedte2018, Ma2021, Bhandari2021, Muechler2022, Chen2023, Gali2023, Haldar2023}. Recent work employing methods that combine quantum chemistry and density functional theory (DFT) via quantum embedding have demonstrated quantitative descriptions of the NV$^{-}$ excited states, at convenient computational costs ~\cite{Bockstedte2018, Ma2021, Muechler2022, Haldar2023}. Thus, developing a quantitative description of strain and charge induced Stark effects on the optical transitions of the NV$^{-}$ through such methods has important implications for NV-based nanoscale sensors.

In this paper, we employ a first-principles approach based on quantum embedding to derive such a description. We study the NV$^{-}$ under strain to derive a many-body description of strain susceptibilities in its optical excited states. We also access the many-body dipole moments of the NV$^{-}$ center and strain induced effects in the optical transitions of the NV center by considering many-body transition dipoles between excited states.

The paper is organized as follows. In Sec.~\ref{sec:methods} we discuss the computational methodology employed in this work and the computational details. Then, we present the obtained results for strain susceptibilities and dipole couplings in Secs.~\ref{sec:A2} and \ref{sec:B2}. This is followed by a discussion of the results in Sec.~\ref{sec:discussion}, and the conclusions  in Sec.~\ref{sec:conclusions}.

\section{Computational methodology \label{sec:methods}}
\subsection{DFT for structure and strain \label{sec:1}}
As a basis for the quantum embedding, we employ DFT calculations to obtain the relaxed atomic structure and the corresponding electronic ground state of a diamond supercell containing a single NV$^{-}$ center, with axis along the $[1,1,1]$. Calculations are performed using the 216-atom (3$\times$3$\times$3) diamond supercell, which is found to yield converged many-body states within our methodology~\cite{Muechler2022}. To study the evolution of the electronic states of the NV$^{-}$ under strain, we introduce distortions to the supercell via direct elongation (contraction) of the lattice vectors for tensile (compressive) strain. For simplicity, we focus on two cases of strain: along the $[1,1,1]$ axis ($\varepsilon_{[ 1, 1, 1 ]}$), and along the $[1,-1,1]$ axis ($\varepsilon_{[ 1, -1, 1 ]}$); see Fig.~\ref{fig:1}{(a)} for a schematic. The second direction of strain, pointing along one of the C--C bonds, is chosen with the aim of breaking the three-fold symmetry of the NV$^{-}$ while best approximating typical strain felt by NV$^{-}$ centers in experiments, i.e., a combination of transverse and longitudinal strains~\cite{Bassett2011, Crisci2011, Acosta2012, Trusheim2016, Lee2016, Happacher2022}. These two directions of strain fields are also experimentally relevant because the NV$^{-}$ is known to respond differently under purely longitudinal or transverse strains ~\cite{Maze2011, Doherty2011, Doherty2012, Gali2019}. The amount of strain applied on each component of the lattice vectors is taken as $\vert\varepsilon\vert$, and defined with respect to the length of the pristine lattice vectors. We will focus on vertical transitions in this work, i.e., neglecting atomic relaxations in the excited state (which would be needed, e.g., to obtain zero-phonon line energies).

\subsection{Quantum embedding \label{sec:2}}
From the ground-state DFT calculation of supercell containing the NV$^{-}$, we downfold onto a minimal active space containing only the defect states of NV$^{-}$ via Wannierization (using the Wannier90 code \cite{Pizzi2020}). The down-folded problem is mapped onto the effective Hamiltonian~\cite{Bockstedte2018, Ma2021, Muechler2022}
\begin{equation}
\label{equation:1}
\begin{split}
\hat{H}_{\text{eff}} & = -\sum_{\braket{ij},\sigma}(t_{ij}c^{\dagger}_{i\sigma}c_{j\sigma} + h.c.) \\
& + \frac{1}{2}\sum_{\braket{ijkl},\sigma, \sigma'} U_{ijkl}c^{\dagger}_{i\sigma}c^{\dagger}_{j\sigma'}c_{l\sigma'}c_{k\sigma} \\ 
& - \hat{H}_{DC},
\end{split}
\end{equation}
where $c^{\dagger}$ ($c$) represent creation (annihilation) operators, and $i$,$j$,$k$,$l$ label different correlated defect states within the active space, with spin $\sigma$, $\sigma'$. 
The first term in Eq.~(\ref{equation:1}) represents the single-particle part of the effective Hamiltonian and is characterized by the hopping matrix elements $t_{ij}$ in the Wannier basis. The hopping matrix elements are determined from the Wannierized basis functions $\phi$ as
\begin{equation}
\label{equation:2}
t_{ij}=-\braket{\phi_{i}|\hat{H}_{\text{KS}}|\phi_{j}}.
\end{equation}
In the absence of disentanglement from bulk bands, the eigenvalues of $t_{ij}$ reproduce the Kohn-Sham (KS) eigenvalues exactly ~\cite{Souza2001}. The second term in Eq.~(\ref{equation:1}) captures the effective electron-electron interactions between correlated states within the active space, in this case represented via the two-body screened Coulomb tensor $U_{ijkl}$. We calculate these effective interactions using the constrained random-phase approximation (cRPA) ~\cite{Aryasetiawan2004}, via
\begin{equation}
\label{equation:3}
U_{ijkl}=\braket{\phi_{i}\phi_{j}|\hat{U}|\phi_{k}\phi_{l}}.
\end{equation}
To do so, we start from the independent-particle polarizability for the full system
\begin{align}
\label{equation:4}
\hat{\Pi}_{\text{full}}(\textbf{r},\textbf{r}',\omega)& = \sum_i^{\text{occ}}\sum_j^{\text{unocc}}\psi_i(\textbf{r})\psi_i^*(\textbf{r}')\psi_j(\textbf{r})\psi_j(\textbf{r}') \\
&\times \left(\frac{1}{\omega-\epsilon_j+\epsilon_i+i\eta} - \frac{1}{\omega-\epsilon_j-\epsilon_i-i\eta}\right)
\end{align}
where $i$ and $j$ are indices of KS bands, $\psi$ and $\epsilon$ are KS wavefunctions and eigenvalues, and $\eta$ is a small positive constant. We will use only the static (i.e., $\omega=0$) macroscopic polarizability in this work, from which we remove the screening processes between occupied and unoccupied NV$^-$ defect states, and obtain a partially-screened Coulomb interaction as
\begin{equation}
\label{equation:5}
\hat{U}=[1-\hat{V}(\hat{\Pi}_{\text{full}}-\hat{\Pi}_{\text{NV}})]\hat{V}.
\end{equation}
Here, ``NV'' refers to the correlated subspace comprised of the defect states associated with the NV$^-$ center, and $\hat{V}$ is the bare Coulomb interaction. In this way, the correlated defect manifold described by Eq.~(\ref{equation:1}) couples to the diamond matrix through the screening effects included while evaluating the Coulomb tensor. Screening effects and correlations within the active space of the defect are included explicitly through the configuration interaction (CI) by diagonalization of Eq.~(\ref{equation:1}) exactly (using the TRIQS software library~\cite{Parcollet2015}). The last term in Eq.~(\ref{equation:1}), so-called double-counting (DC) correction, is added to deal with unwanted Coulomb interactions partially included (from DFT) in the single-particle parts of the full effective Hamiltonian. Here, we adopt a DC correction of the form~\cite{Muechler2022, Ma2021, Bockstedte2018}
\begin{equation}
\label{equation:6}
\hat{H}_{DC}=\sum_{\braket{ij},\sigma}c^{\dagger}_{i\sigma}c_{j\sigma}\sum_{\braket{kl}}P_{kl}(U_{iljk}- \alpha U_{ilkj}),
\end{equation}
where, following Ref.~\cite{Bockstedte2018}, $\alpha$ relates to the amount of Hartree-Fock exchange mixed to construct the DFT functional. Since we use the Heyd-Scuseria-Ernzerhof (HSE) hybrid functional in all DFT calculations, we set $\alpha = 0.25$.
Employing this DC scheme has shown to yield NV$^{-}$ excitation energies in good agreement with those from experiments \cite{Bockstedte2018, Muechler2022}. In any case, the quantities of interest in this work and their trends with strain are quantitatively very similar to those obtained without DC corrections.

\subsection{Many-body dipole couplings \label{sec:3}}
To describe the dipole couplings to electric fields of the NV$^{-}$, we evaluate the dipole matrix elements across the many-body spectrum via
\begin{equation}
\label{equation:7}
\bm{\mu}_{ij}=\braket{\Psi_{i}|{\hat{\textbf{r}}}|{\Psi_{j}}},
\end{equation}
where, $\Psi$ denotes many-body wave functions of states $i$ and $j$ (respectively), and $\hat{\textbf{r}}$ represents the many-body dipole operator (along the Cartesian directions). In our formalism, we construct the many-body dipole operator as
\begin{equation}
\label{equation:8}
{\hat{\textbf{r}}}=\sum_{nm}\textbf{r}^{\text{wann}}_{nm}({\bm 0}) c^{\dagger}_{n}c_{m},
\end{equation}
using the matrix elements of the position operator between Wannier functions $m$ and $n$, given by
\begin{equation}
\label{equation:9}
\textbf{r}^{\text{wan}}_{nm}(\textbf{R})=\braket{n\textbf{0}|\textbf{r}|m\textbf{R}},
\end{equation}
as implemented in Wannier90 ~\cite{Pizzi2020}. This approach allows for permanent and transition dipole matrix elements between explicitly \emph{many-body} ground and excited states of NV$^{-}$. We note that some properties of the dipole matrix elements might depend on the specific gauge chosen for constructing our Wannier functions. To quantify this, we evaluated the dipole matrix (in the absence of strain) using a Wannierization procedure slightly different from that employed for the main results (Sec.~\ref{sec:4}), i.e., increasing the disentanglement window slightly, but other properties of our many-body model (i.e., energies, spin state, symmetries) are not significantly altered.

\subsection{Wannierization \label{sec:4}}
In terms of the Wannierization scheme specifically implemented for the NV$^{-}$ defect center, we rely on initial projections of $sp^{3}$-hybrid Wannier functions centered around the NV$^{-}$ to obtain the localized atomic basis orbitals. This basis is utilized to evaluate the screened Coulomb interactions and subsequently perform all cRPA-CI calculations. The single-particle electronic states relevant for most of the optical excitations of the NV$^{-}$ center are the $a_{1}(1)$, $a_{1}(2)$, $e_{x}$, and $e_{y}$. While the $a_{1}(2)$, $e_{x}$, and $e_{y}$ lie isolated within the bandgap, the $a_{1}(1)$ state is resonant with the valence band (VB) of diamond, requiring disentanglement during the Wannierization procedure. Hence, we set our disentanglement window to include the $a_{1}(1)$ state, while the $a_{1}(2)$, $e_{x}$ and $e_{y}$ states are kept ``frozen'' such that Eq.~(\ref{equation:2}) reproduces the DFT eigenvalues exactly. All Wannierized states are maximally-localized by minimizing the Wannier spreads. Overall, our active space is comprised of the four $a_{1}(1)$, $a_{1}(2)$, $e_{x}$, and $e_{y}$ Wannierized orbitals and 6 electrons. This active space has demonstrated to be sufficient to yield quantitative agreement between calculated and experimental excitation energies for the NV$^{-}$ ~\cite{Bockstedte2018, Muechler2022}. The main task concerning our current work centers on studying how this active space and its many-body states evolve under the external influence of strain, as we discuss in Sec.~\ref{sec:results}.

\begin{figure*}[h! t]
\centering
\includegraphics[width=1.0\linewidth]{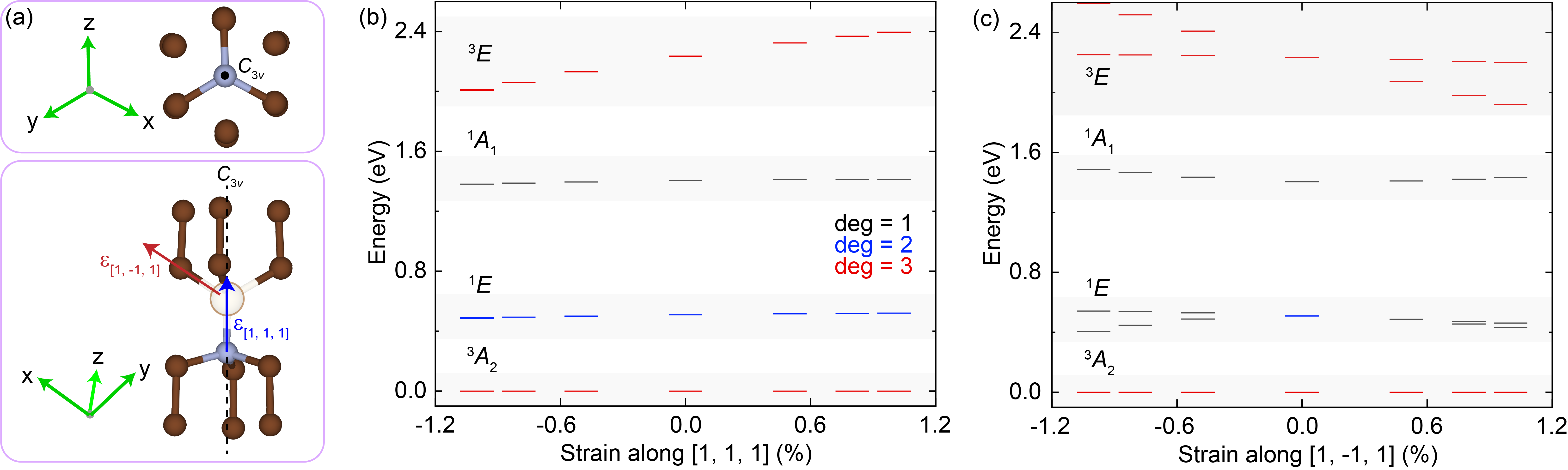}
\caption{Many-body spectrum of the NV$^{-}$ center in the presence of strain. (a) Simplified atomic structure of the NV center in diamond. The high-symmetry and directions of applied strain are illustrated for clarity. Carbon and nitrogen atoms are represented by brown and grey spheres, respectively, while the carbon vacancy is depicted with the light brown circle. (b) $\varepsilon_{[1,1,1]}$ introduces energy shifts of the $^{3}E$ manifold with respect to the $^{3}A_{2}$ ground state. Energy shifts of the other manifolds are comparatively much smaller. (c) $\varepsilon_{[1,-1,1]}$ introduces an energy split of the $^{3}E$ into the $^{3}E_{x/y}$ sub-manifolds. The $^{1}E$ manifold also splits under such strain, while the $^{1}A_{1}$ state shows larger shifts compared to the $[1,1,1]$-strained case.}
\label{fig:1}
\end{figure*}

\section{Results \label{sec:results}}

\subsection{Strain susceptibilities of NV$^{-}$ \label{sec:A2}}
We start by considering the case of strain along the $[1,1,1]$. Since this strain occurs along the high-symmetry axis of the NV$^-$ center ($A_{1}$ strain), the $C_{3v}$-symmetry of the defect is preserved [Fig.~\ref{fig:1}{(a)}]. As a result, the $^{3}E$ states remain degenerate, but shift in energy with respect to the ${^2}A_{2}$ ground state as a function of $\varepsilon_{[1,1,1]}$ [Fig.~\ref{fig:1}{(b)}]. The lack of inversion symmetry in the NV center is reflected in the asymmetric evolution of the excited states under $\pm$ $\varepsilon$ (elongation or compression). Overall, these results indicate that in the scenario where local strain is only longitudinal to the NV$^{-}$ axis, a red-shift in the optical transition of the NV$^{-}$ corresponds to local compressive strain, while tensile strain will manifest as a blue shift in the optical transition. Lastly, Fig.~\ref{fig:1}{(b)} clearly illustrates that the energy shifts are much larger for the $^{3}E$ manifold compared to the singlet excited states, confirming that the triplet manifold is much more sensitive to longitudinal strain.

Under $\varepsilon_{[1,-1,1]}$, the picture is more complicated. Here, the strain has both longitudinal and transverse components to the local NV$^-$ axis [red arrow in Fig.~\ref{fig:1}{(a)}]. The $C_{3v}$ symmetry is thus broken and the degeneracy of the $E$ states is lifted [Fig.~\ref{fig:1}{(c)}] \cite{Maze2011}. The evolution of the excited states under $\varepsilon_{[1,-1,1]}$ still includes energy shifts caused by the longitudinal strain components. Again, the $^{3}E$ states show the largest strain induced effects. However, in this case the $^{1}E$ states also display a relatively strong response. This splitting correlates well with observations of split optically-detected magnetic resonance through the singlet transition ~\cite{Rogers2008, Acosta2010}. Thus, the singlet manifold could be an interesting probe for strain sensing under specific conditions, e.g., in the near-infrared spectral region, given its weak/strong response to longitudinal/transverse strains [Figs.~\ref{fig:1}{(b)} and \ref{fig:1}{(c)}]. However, the inter-system crossing required for populating the singlet manifold is typically a slow process, which, combined with a smaller radiative efficiency and longer decay lifetime ~\cite{Rogers2008, Acosta2010}, makes the photon collection in the singlet manifold poorer compared to the triplet optical transitions. With this in mind, we make the $^{3}E$ manifold the focus of the remainder of this paper.

\begin{figure*}[h t]
\centering
\includegraphics[width=1.0\linewidth]{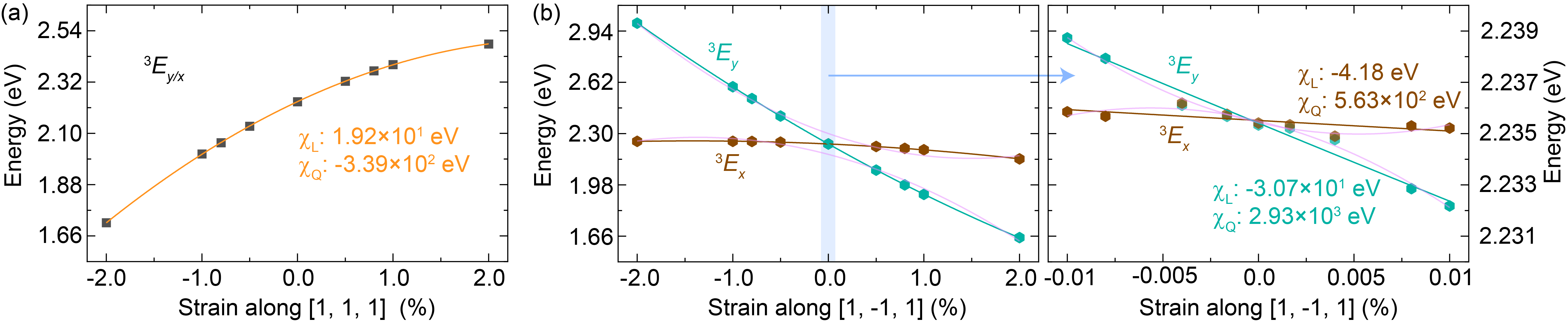}
\caption{Strain susceptibility ($\chi$) for the $^{3}E$ many-body states of the NV$^{-}$ center. (a) $^{3}E_{y/x}$ under $\varepsilon_{[1,1,1]}$, covering a wide range of strain. (b) Left panel: Same as in (a), but for the case of $\varepsilon_{[1,-1,1]}$. A set of higher--resolution calculations for this case of strain (within violet-shaded region) is included in the right panel of (b), to better resolve the response within the strain region most relevant to experiments ($\varepsilon < 10^{-5}$). The scatter and solid lines represent numerical data and second-degree polynomial fits, respectively.}
\label{fig:2}
\end{figure*}

In Fig.~\ref{fig:2}, we make use of the strain evolution of the $^{3}E$ states to extract the strain susceptibility (denoted here as $\chi$) of the optical transitions in the NV center. As mentioned previously, $\varepsilon_{[1,1,1]}$ does not split the $^{3}E$ manifolds. In this case, $\chi$ is identical for the $^{3}E_{y/x}$ states [Fig.~\ref{fig:2}{(a)}], and is found to follow an almost-linear response, with linear coefficient $\chi_{\text{L}}=19.2$ eV. The wide range of strain considered here allows to also capture the quadratic component of the strain susceptibility $\chi_{\text{Q}}$, although the realistic range of strain (typically $< 10^{-5}$) renders this contribution negligible within linear-strain regimes most relevant for experiments.

Because transverse strain breaks the degeneracy between $^{3}E_{y/x}$ states [left panel of Fig.~\ref{fig:2}{(b)}], care must be taken in how the states are labeled. For instance, if one assumes that the energy ordering stays the same for different signs of strain, e.g., $E_{y}$ is always the lowest in energy, then in the case of $\varepsilon_{[1,-1,1]}$ there is an apparent anti-crossing, and the quadratic components of the susceptibility over the same range of strain as in Fig.~\ref{fig:2}{(a)} are more pronounced [illustrated by pink fits to the data in left panel of Fig~\ref{fig:2}{(b)}]. This is consistent with what has been done previously in the experimental literature \cite{Tamarat2006, Bassett2011, Acosta2012, Lee2016, McCullian2022}.

On the other hand, labeling the $E_{y/x}$ states based on their strain induced shifts, e.g., $E_{y}$ as the state with largest shift with strain, we find a much more linear response for both $E_{y}$ and $E_{x}$ states [solid fits in Fig.~\ref{fig:2}{(b)}]. Further, the strongest susceptibility of the $E_{y}$ in this picture is consistent with the transverse component of strain being primarily along the local $y$ direction [Fig.~\ref{fig:1}{(a)}]. Lastly, as shown in the next section, this labeling of the states results in a smooth evolution of the dipole couplings, with clear trends under the effect of strain.

The magnitude of strain fields experienced by NV$^-$ centers in typical experiments is well below the strains considered up to this point, usually below $\varepsilon=\pm 10^{-5}$ ~\cite{Bassett2011, Crisci2011, Acosta2012, Trusheim2016, Lee2016, Happacher2022, Ji2024, Delord2024, McCullian2024}. Again, within this range of strain, the responses to $\varepsilon_{[1,1,1]}$ for the $^{3}E_{y/x}$ states should be effectively linear. Thus, to quantify $\chi_{\text{L}}$ more reliably for transverse strain, we expand the resolution by considering a narrower range of strains around zero ($\varepsilon < 10^{-4}$) for $\varepsilon_{[1,-1,1]}$ [highlighted region in Fig.~\ref{fig:2}{(b)}, left panel]. The obtained results for these calculations are presented in the right panel of Fig.~\ref{fig:2}(b). The fit to the numerical data obtained in this case is also improved by exclusion of the extreme points $\varepsilon > 1.0 \%$, which are in any case well outside any realistic experimental situation. Compared to Fig.~\ref{fig:2}{(b), left panel}, the set of points around zero-strain resemble a clearer linear response and the $\chi_{\text{L}}$ extracted from this fit should be closer to that expected for calculations in the linear regime of strain. 

Ideally, one would want to consider even weaker strain, such that the calculated strain evolution of the $^{3}E_{y/x}$ states is explicitly matched to that obtained experimentally. However, such strain fields require very small atomic displacements, as well as the resolution of correspondingly small energy splitting of the single-particle/many-body states, which are close to the numerical precision of our methods. This is evident from the larger scatter present for the points with $|\varepsilon_{[1,1,1]}| < 0.005$~\% in Fig.~\ref{fig:2}{(b), right panel}, which mainly originates from a lack of resolution in the single-particle energy splitting within this range of strain due to finite electronic smearing in the DFT calculations.

\subsection{Dipole couplings of NV$^{-}$ under strain \label{sec:B2}}
\begin{table*}[ht]
\setlength{\tabcolsep}{1pt}
\renewcommand{\arraystretch}{1.5}
\caption{Dipole couplings for the NV$^{-}$ under strain. The format of the dipole couplings presented below as vector components in parentheses followed by the norm. In all cases, the applied strain is along the $[1,-1,1]$ direction (negative values correspond to compressive strain).}
\label{table:1}
\centering
\begin{tabularx}{\textwidth}{c*{4}{>{\centering\arraybackslash}X}}
\hline\hline
$\varepsilon_{[1,-1,1]}$ (\%) & ($\Delta\textbf{p}_y$), $\vert\Delta\textbf{p}_y\vert$ & ($\Delta\textbf{p}_x$), $\vert\Delta\textbf{p}_x\vert$ & ($\Delta\textbf{p}_{\text{avg}}$), $\vert\textbf{p}_{\text{avg}}\vert$& ($\bm{\mu}_{xy}$), $\vert\bm{\mu}_{xy}\vert$\\
\hline\hline
$-0.01$ & (1.83, $-0.82$, 1.81), 2.70 & (0.06, 2.71, 0.08), 2.71 & (0.95, 0.95, 0.95), 1.65 & ($-1.32+0.76i$, $-0.03-0.03i$, $1.34-0.74i$), 2.16 \\
\hline
$-0.008$ & (1.84, $-0.82$, 1.81), 2.71 & (0.05, 2.71, 0.08), 2.71 & (0.95, 0.95, 0.95), 1.65 & ($1.13+1.01i$, $0.02$, $-1.15-1.01i$), 2.15 \\
\hline
0.00 & (1.70, 1.68, $-0.55$), 2.45 & (0.19, 0.21, 2.43), 2.45 & (0.95, 0.95, 0.94), 1.64 & --
\\
\hline
+0.008 & (1.81, $-0.82$, 1.83), 2.70 & (0.07, 2.71, 0.05), 2.71 & (0.94, 0.95, 0.94), 1.63 & ($-1.53+0.02i$, $0.01-0.02i$, 1.52), 2.16 \\
\hline
+0.01 & (1.82, $-0.82$, 1.83), 2.71 & (0.07, 2.71, 0.05), 2.71 & (0.95, 0.95, 0.94), 1.63 & ($-0.79-1.31i$, $0.02$, $0.78+1.31i$), 2.16 \\
\hline\hline
\end{tabularx}
\end{table*}

Within our methodology, we can obtain a many-body description of the permanent dipole and transition dipole moments of the NV$^{-}$, as well as possible strain induced effects on these dipole couplings. The diagonal elements of the full dipole matrix described in Sec.~\ref{sec:3} [Eq.~(\ref{equation:7})] are used to extract the permanent dipoles, $\textbf{p}_i\equiv{\bm \mu}_{ii}$. We take the many-body ground-state dipole as reference~\cite{Gali2019}, i.e., $\Delta\textbf{p}_{y/x} \equiv \braket{^{3}E_{y/x}|\hat{\textbf{r}}|^{3}E_{y/x}}-\braket{^{3}A_{2}|\hat{\textbf{r}}|^{3}A_{2}}$ for transitions to individual $^3E$ states. However these values are only relevant if experimental conditions allow to resolve the $E_x$ and $E_y$ states separately; if the temperature is too high compared to the strain-induced splitting of these levels, or off-resonant excitation is used, the relevent quantity is the vector average of the two perminant dipoles, which we denote $\Delta\textbf{p}_{\text{avg}} = (\braket{^{3}E_y|\hat{\textbf{r}}|^{3}E_y}+\braket{^{3}E_x|\hat{\textbf{r}}|^{3}E_x})/2-\braket{^{3}A_{2}|\hat{\textbf{r}}|^{3}A_{2}} \equiv \braket{^{3}E|\hat{\textbf{r}}|^{3}E}-\braket{^{3}A_{2}|\hat{\textbf{r}}|^{3}A_{2}}$. The directions and magnitudes of the dipole couplings for a few values of $
\varepsilon_{[1,-1,1]}$ are provided in Table ~\ref{table:1}.

In the absence of strain, the individual permanent dipoles have magnitude of $\vert\Delta\textbf{p}_{y/x}\vert=2.4$ D, while $\vert\Delta\textbf{p}_{\text{avg}}\vert = 1.6$ D; upon the application of transverse strain and the breaking of the degeneracy between the levels, $\vert\Delta\textbf{p}_{y/x}\vert$ increases slightly to 2.7 D. Therefore, we see that an optical  measurement that can resolve $E_x$ and $E_y$ would be more sensitive to electric fields, with an enhancement factor of $\vert\Delta\textbf{p}\vert$ is $\sim 1.5$. Also, such a measurement would have more comprehensive directional sensitivity, since the individual dipoles have projections in other directions than just the NV$^{-}$ axis.

We can understand the relationships between the directions and magnitudes of the permanent and transition dipole moments via the symmetry analysis described by Maze \textit{et al.} \cite{Maze2011}. Firstly, they pointed out that due to the three-fold rotational axis, $^3A_2$ only couples to longitudinal fields (i.e., along the axis connecting N and the vacancy, denoted here as NV$^{-}_z$), with a different coupling parameter to $^3E$, while $^3E$ couples to both fields longitudinal and transverse to NV$^{-}_z$. The sign of the coupling for $E_x$ and $E_y$ to the transverse fields is opposite, and, interestingly, the transition dipole matrix element $\bm{\mu}_{xy}\equiv\braket{^{3}E_{y}|{\hat{\textbf{r}}}|^{3}E_{x}}$ should have the same magnitude as the transverse permanent dipole $\vert\textbf{p}_{x/y}\vert$ ~\cite{Maze2011}.

We start with the zero strain case, where $E_x$ and $E_y$ are degenerate and the NV$^-$ has $C_{3v}$ point symmetry. The vectors $\Delta\textbf{p}_{x/y}$ include a component along NV$^{-}_z=[1,1,1]$ [see Fig.~\ref{fig:1}(a)], which is the difference of longitudinal-field coupling parameters between $^3A_2$ and $^3E$. This value can be obtained either by projecting either $\Delta\textbf{p}_{x}$ or $\Delta\textbf{p}_{y}$ along [1,1,1], or, simply averaging the two since the transverse-field coefficients have opposite sign and cancel. Thus  $\Delta\textbf{p}_{\text{avg}}$ is in [1,1,1] as we see in Table~\ref{table:1} (fourth column). Removing this longitudinal component gives $\Delta\textbf{p}_{y}-\Delta\textbf{p}_{\text{avg}}\propto [1,1,-2]$ and $\Delta\textbf{p}_{x}-\Delta\textbf{p}_{\text{avg}}\propto [-1,-1,2]$, though we expect the specific dipole directions of these degenerate states to depend on the basis chosen to represent them.

Now we consider the cases in Table~\ref{table:1} corresponding to finite strain, where the symmetry is lowered to $C_s$, i.e., the three-fold symmetry is broken and only one mirror plane remains. We see that for the (small) strains shown, $\Delta\textbf{p}_{\text{avg}}\propto[1,1,1]$, indicating that similar arguments apply as for the no-strain $C_{3v}$ case, even though the three-fold symmetry is broken and the $^3A_2$ dipole is not necessarily constrained to [1,1,1] (which may be the reason for the small deviations seen at positive strains). Removing the [1,1,1] component for the finite strain cases gives $\Delta\textbf{p}_{y}-\Delta\textbf{p}_{\text{avg}}\propto [1,-2,1]$ and $\Delta\textbf{p}_{x}-\Delta\textbf{p}_{\text{avg}}\propto [-1,2,-1]$. This is consistent with the symmetry requirement that the permanent dipoles lie in the single mirror plane that remains after $\varepsilon_{[1,-1,1]}$ is applied. (This plane contains both $[1,1,1]$ and $[1,-1,1]$, and thus is perpendicular to $[-1,0,1]$.) Also, $\vert\Delta\textbf{p}_{x/y}-\Delta\textbf{p}_{\text{avg}}\vert=2.16$ D consistent with the calculated $\vert\bm{\mu}_{xy}\vert$ in the last column of Table~\ref{table:1} \cite{Maze2011}. In contrast, the \emph{direction} of the transition dipole moment  $\bm{\mu}_{xy}$ must have (at least a component) \emph{perpendicular} to the remaining mirror plane. This is because when transverse strain lowers the symmetry of the NV$^-$ from $C_{3v}$ to $C_s$, it splits states transforming like the $E$ irreducible representation into $A'$ and $A''$; the former is even under the mirror symmetry, and the later is odd, so to cause a transition the light polarization must have a mirror-odd component, i.e., a dipole perpendicular to the mirror plane. We see from the last column in Table~\ref{table:1} that this is exactly the case, with both real and imaginary components of $\bm{\mu}_{xy}$ pointing in $[-1,0,1]$ (or $[1,0,-1]$). In addition to fulfilling the expected symmetry considerations, the dipole directions and magnitudes correlate well with experimental observations; in particular, the magnitudes of both $\Delta\textbf{p}$ and $\bm{\mu}_{xy}$ fall well within experimental estimations, while their relative orientations align with the direction of local strain~\cite{Tamarat2006, Mittiga2018, Bian2021, Block2021, Ji2024, Delord2024}.

Other effects, such as phonon scattering and local strain environment, will likely impact the effective enhancement of the $\vert\Delta\textbf{p}\vert$ under resonant excitation (mentioned above). In fact, by considering strain effects explicitly, we find that $\vert\Delta\textbf{p}_{y/x}\vert$ increases for $\varepsilon_{[1,-1,1]}$, but decreases with $\varepsilon_{[1,1,1]}$ (not shown). Naturally, the amount of enhancement/suppression of the permanent dipole depends on the magnitude of local applied strain, but from the results in Table ~\ref{table:1}, it corresponds to $< 12.5 \%$ for $\vert\varepsilon\vert \leq 0.008 \%$. Since $\varepsilon_{[1,-1,1]}$ is not fully perpendicular to the local NV$^-_z$ axis, and introduces a transverse strain component, these results suggest that $\varepsilon \perp$ NV$^{-}_{z}$ could slightly enhance $\Delta\textbf{p}_{y/x}$ ~\cite{Delord2024-2}. Additionally, the average permanent dipole remains roughly constant with strain, which means that the relative enhancement of the $\Delta\textbf{p}_{y}$ components with strain is anti-correlated with that of $\Delta\textbf{p}_{x}$, and if at all present, will only be quantifiable via resonant-excitation experiments at cryogenic temperatures. In terms of directionality, the obtained results show that while the individual dipoles change directions with $\varepsilon_{[1,-1,1]}$, they always maintain anti-parallel alignment of their off-axis component, such that $\Delta\textbf{p}_{\text{avg}}$ is always along the $[1,1,1]$ (Table ~\ref{table:1}). It is worth noting that the $\vert\Delta\textbf{p}_{\text{avg}}\vert$ magnitudes obtained herein are in excellent agreement with experimental estimations ~\cite{Tamarat2006, Ji2024} (similar to that for the NV$^{0}$~\cite{Kurokawa2024}), also falling close to those derived through DFT ~\cite{Gali2019, Alaerts2024}.

\section{Discussion \label{sec:discussion}}
As previously mentioned, there is a decently wide spread on the permanent dipoles for the NV$^{-}$ center calculated from first principles. Hence, it is important to put our results in the context of previous theoretical work. Firstly, employing the same methodology described in Sec.~\ref{sec:3}, but disregarding interactions within the embedding Hamiltonian (single-particle picture of states and dipole moments), we obtain a net permanent dipole of 3.01 D in the absence of strain. This number, in decent agreement with that derived from DFT using finite fields ~\cite{Alaerts2024}, represents the permanent dipole from wave functions constructed using our single-particle occupation and basis states. Compared to our result of $\vert\Delta\textbf{p}_{\text{avg}}\vert = 1.63$ D, we see that the many-body effects renormalize the dipoles by almost a factor of two. We note that the discrepancy between our single-particle permanent dipole and the 2.4 D of Ref.~\cite{Alaerts2024} may be due to additional effects related to the field polarizing the surrounding diamond, which are not included within our methodology. Additionally, the difference between our single-particle permanent dipole and the 4.3 D directly derived from the Berry-phase polarization approach in DFT, e.g., as reported in~\cite{Gali2019} and~\cite{Alaerts2024}, suggests that there could be an issue with the implementation of dipole matrix element calculations at the DFT level with constrained occupation of the excited state (also mentioned in Ref.~\cite{Alaerts2024}). Lastly, the 0.79 D reported in Ref.~\cite{Maze2011} makes use of a different approach; they assume that the corresponding Stark effect is due to piezoelectric coupling with the atoms that make up the NV$^{-}$ defect. In all of our work (and that of Refs.~\cite{Gali2019, Alaerts2024}) the atoms are kept fixed, which should explain the discrepancy. We leave the calculation of the combined atomic and electronic effects to future work.

Overall, the results described in Sec.~\ref{sec:results} provide useful insights on the intricate response of the NV$^{-}$ to local strain and electric fields. Firstly, the asymmetric evolution of the excited states with strain appears to be more prominent when the strain has both transverse and parallel components to the NV$_{\text{z}}$ axis. This behavior, emerging from the absence of inversion symmetry on the NV$^{-}$ center, should provide a means of differentiating local compressive versus tensile strains. On the other hand, from the relative energy shifts and splitting under $\varepsilon_{[1,1,1]}$ and $\varepsilon_{[1,-1,1]}$ shown in Fig.~\ref{fig:1}, one could argue that the singlet transition is more reliable at differentiating transverse versus longitudinal strains than the triplet transition. This could be an interesting application of the singlet transitions in the NV$^{-}$ for local strain sensing: the triplet manifold reacts to both longitudinal and transverse strains (via shifts and splitting of the optical line, respectively), while the singlet reacts (splits) most prominently only to transverse components of strain. In essence, correlated high-resolution measurements in which both the triplet and singlet transitions are interrogated through resonant excitations should provide useful information about strain propagation through diamond at the nanoscale.

Along these lines, our results hint at the $^{3}E_{y} \leftrightarrow ^{3}E_{x}$ transition as an additional asset for strain sensing using the NV$^{-}$. Since this transition appears almost fully linearly polarized, and its directionality reacts to the direction of the overall strain, it could help decompose the local strain in its vector components. In principle, the measurements required to probe $\bm{\mu}_{yx}$ are independent from those required to monitor the Stark shifts on the ZPL through the permanent dipole. Thus, combining information from these two complimentary dipole couplings should provide an enhanced probe for strain sensing using the NV$^-$ center. Further, given the increasing interest in studying closely-positioned NV$^-$ centers in diamond ~\cite{Lozovoi2021, Ji2024, Delord2024, Delord2024-2}, the proposed measurements could serve as useful probes for possible NV--NV interactions mediated via strain.

\section{Conclusions \label{sec:conclusions}}
We have employed first-principles calculations based on density-functional theory and quantum embedding to derive a many-body description of strain and charge induced Stark effects in the NV$^{-}$ defect center in diamond. The relaxed atomic structures of a diamond supercell containing a single NV$^{-}$ under the influence of strain have been obtained from DFT. Using these as starting points for embedding calculations, we extract the strain susceptibility of the many-body states comprising the $^{3}E$ excited-state manifold, as well as permanent and transition dipoles under strain. The results derived in this work are in excellent agreement with recent experiments, where the dipole couplings of the NV$^{-}$ are estimated and/or used for sensing applications ~\cite{Tamarat2006, Ji2024, Delord2024, McCullian2024, Delord2024-2}. This work thus broadens our fundamental understanding on the optoelectronic properties of the NV$^{-}$ center, and also highlights the utility of including many-body effects to describe the properties of defect-related color centers in solids for sensing applications. 

\acknowledgements
CED acknowledges support from the National Science Foundation under Grant No.~DMR-2237674. GILM acknowledges funding from grant NSF-2208863. JZ and CED acknowledge support from a joint SBU-BNL seed grant. CAM acknowledges support by the U.S. Department of Energy, Office of Science, National Quantum Information Science Research Centers, Co-design Center for Quantum Advantage (C2QA) under contract number DE-SC0012704. The Flatiron Institute is a division of the Simons Foundation.

\bibliography{NV_bib}

\end{document}